# Protein Contact Prediction by Integrating Joint Evolutionary Coupling Analysis and Supervised Learning

**Jianzhu Ma**   **Sheng Wang**   **Zhiyong Wang**   **Jinbo Xu**

Toyota Technological Institute at Chicago
{*majianzhu, wangsheng, zywang, j3xu*}@ttic.edu

## Abstract

Protein contacts contain important information for protein structure and functional study, but contact prediction from sequence remains very challenging. Both evolutionary coupling (EC) analysis and supervised machine learning methods are developed to predict contacts, making use of different types of information, respectively. This paper presents a group graphical lasso (GGL) method for contact prediction that integrates joint multi-family EC analysis and supervised learning. Different from existing single-family EC analysis that uses residue co-evolution information in only the target protein family, our joint EC analysis uses residue co-evolution in both the target family and its related families, which may have divergent sequences but similar folds. To implement joint EC analysis, we model a set of related protein families using Gaussian graphical models (GGM) and then co-estimate their precision matrices by maximum-likelihood, subject to the constraint that the precision matrices shall share similar residue co-evolution patterns. To further improve the accuracy of the estimated precision matrices, we employ a supervised learning method to predict contact probability from a variety of evolutionary and non-evolutionary information and then incorporate the predicted probability as prior into our GGL framework. Experiments show that our method can predict contacts much more accurately than existing methods, and that our method performs better on both conserved and family-specific contacts.

## Introduction

Protein contacts contain important information for protein folding and recent works indicate that one correct long-range contact for every 12 residues in the protein allows accurate topology level modeling (Kim, et al., 2013). Thanks to high-throughput sequencing and better statistical and optimization techniques, evolutionary coupling (EC) analysis for contact prediction has made good progress, which makes de novo prediction of some large proteins possible (Hopf, et al., 2012; Marks, et al., 2011; Nugent and Jones, 2012; Sułkowska, et al., 2012). Nevertheless, contact prediction accuracy is still low even if only the top L/10 (L is the sequence length) predicted contacts are evaluated.

Existing contact prediction methods can be roughly divided into two categories: 1) evolutionary coupling (EC) analysis methods, such as (Burger and van Nimwegen, 2010; Di Lena, et al., 2011; Marks, et al., 2011), that make use of multiple sequence alignment; and 2) supervised machine learning methods, such as SVMSEQ (Wu and Zhang, 2008), NNcon (Tegge, et al., 2009), SVMcon (Cheng and Baldi, 2007), CMAPpro (Di Lena, et al., 2012), that predict contacts from a variety of information including mutual information, sequence profile and some predicted structure information. In addition, a couple of methods also use physical constraints, such as PhyCMAP (Wang and Xu, 2013) and Astro-Fold (Klepeis and Floudas, 2003).

Residue EC analysis is a pure sequence-based method that predicts contacts by detecting co-evolved residues from the MSA (multiple sequence alignment) of a single protein family. This is based upon an observation that a pair of co-evolved residues is often found to be spatially close in the 3D structure. Mutual information (MI) is a local statistical method used to measure residue co-evolution strength, but it cannot tell apart direct and indirect residue interaction and thus, has low prediction accuracy. Along with many more sequences are available, some global statistical methods, such as maximum entropy and probabilistic graphical models, are developed to infer residue co-evolution from MSA (Balakrishnan, et al., 2011; Cocco, et al., 2013; Jones, et al., 2012; Lapedes, et al., 2012; Lapedes, et al., 1999; Marks, et al., 2011; Thomas, et al., 2008; Thomas, et al., 2009; Thomas, et al., 2009; Weigt, et al., 2009). These global statistical methods can differentiate direct from indirect residue couplings and thus, are more accurate than MI. See (de Juan, et al., 2013) for an excellent review of EC analysis. Representative tools of EC analysis include Evfold (Marks, et al., 2011), PSICOV (Jones, et al., 2012), GREMLIN (Balakrishnan, et al., 2011), and plmDCA (Ekeberg, et al., 2013). Meanwhile, GREMLIN and plmDCA do not assume Gaussian distribution of a protein family. All these EC methods make use of residue co-evolution information only in the target protein family, ignoring other related families.

Supervised machine learning methods (Cheng and Baldi, 2007; Shackelford and Karplus, 2007; Wang and Xu, 2013) make use of not only mutual information (MI), but also sequence profile and other protein features, as opposed to EC analysis that makes use of only residue co-evolution. Experiments show that due to use of more information, supervised learning may outperform EC methods especially for proteins with few sequence homologs (Wang and Xu, 2013). Recently, a few groups such as DNcon (Eickholt and Cheng, 2012), CMAPpro (Di Lena, et al., 2012) and PConsC (Skwark, et al., 2013) have also applied deep learning, an emerging supervised learning method, to contact prediction and showed some improved performance.

In this paper, we present a new method CoinDCA (co-estimation of inverse matrices for direct-coupling analysis) for contact prediction that integrates joint multi-family EC analysis and supervised machine learning. Since joint EC analysis and supervised learning use different types of information, their combination shall lead to better prediction accuracy. The contribution of this paper lies in two aspects. First, different from existing EC analysis that makes use of residue co-evolution information only in the target protein family, our joint EC analysis predicts contacts of a target family not only using its own residue co-evolution information, but also those in its related families which may share similar contact maps. By enforcing contact map consistency in joint EC analysis, we can greatly improve contact prediction. To fulfill this, we develop a statistical method called group graphical lasso (GGL) to estimate the joint probability distribution of a set of related families and enforce contact map consistency proportional to evolutionary distance. Second, we use Random Forests, a popular supervised learning method, to predict the probability of two residues forming a contact using a variety of evolutionary and non-evolutionary information. Then we integrate the predicted probability as prior into our GGL formulation to further improve the accuracy of joint EC analysis.

Experiments show that our method greatly outperforms existing EC or supervised machine learning methods regardless of the number of sequence homologs available for a target protein under prediction, and that our method not only performs better on conserved contacts, but also on family-specific contacts. We also find out that contact prediction may be worsened by merging multiple related families into a single one followed by single-family EC analysis, or by consensus of single-family EC analysis results.

## Results and Discussion

**Test Data.** It is selected from the 150 Pfam (Bateman, et al., 2004; Finn, et al., 2014) families used by PSICOV (Jones, et al., 2012) as benchmark, all of which have solved structures in PDB. To make a fair comparison, we use the same solved structures as PSICOV to calculate native contacts. In this paper we consider only $C_\alpha$ contacts and similar performance trend is observed for $C_\beta$ contacts. We denote these Pfam families, for which we would like to predict contacts, as the target families. For each target family, we find its related families in Pfam, also called auxiliary families, using HHpred (Söding, 2005) with E-value=$10^{-6}$ as cutoff. As a result, 98 families have at least one auxiliary family and are used as our test data. We can also relax the E-value cutoff to obtain more distantly-related auxiliary families, but this does not lead to significantly improved prediction accuracy. Among the 98 target families, the average TMscore (Zhang and Skolnick, 2005) between the representative solved structures of a target family and of its auxiliary families is ~0.7. That is, on average the target and auxiliary families are not very close, although they may have similar folds. Even using E-value$\leq 10^{-17}$ as cutoff, some target and auxiliary families are only similar at the SCOP fold level.

**Evaluation**. To save space, we tested our method against a few popular EC methods such as PSICOV, Evfold, plmDCA (Ekeberg, et al., 2013) and GREMLIN (Kamisetty, et al., 2013) and a few supervised learning methods such that NNcon and CMAPpro, all of which were run with default parameters.

To ensure that the Pfam database does not miss important sequence homologs, we also generate an MSA for each target family by PSI-BLAST (5 iterations and E-value=0.001) and then apply PSICOV to this MSA. Such a method is denoted as PSICOV_b. Since HHblits (Remmert, et al., 2012) sometimes may detect sequence homologs of higher quality than PSI-BLAST, we also run HHblits to build MSA for a target sequence and then examine if HHblits-generated MSA can lead to better prediction for plmDCA and GREMLIN or not.

There are two alternative strategies to use information in auxiliary families. One is that we can merge a target and its auxiliary families into a single one and then apply single-family EC analysis. To test if this strategy works or not, we align and merge a target and its auxiliary families into a single MSA using a probabilistic consistency method (Do, et al., 2006; Peng and Xu, 2011) and MCoffee (Wallace, et al., 2006), respectively. We denote them as Merge_p and Merge_m, respectively. The other strategy is that we apply the single-family EC method PSICOV to each of the target and auxiliary families and then apply a majority voting method to predict the contacts in the target family. We denote this strategy as Voting.

We evaluate the top L/10, L/5 and L/2 predicted contacts where L is the sequence length of a protein (family) under prediction. When more predicted contacts are evaluated, the difference among methods decreases since it is more likely to pick a native contact by chance. Contacts are short-, medium- and long-range when the sequence distance between the two residues in a contact falls into three intervals [6, 12], (12, 24], and >24, respectively. Generally speaking, medium- and long-range contacts are more important, but more challenging to predict.

## Overall performance on the PSICOV test set

As shown in Table 1, tested on all the 98 Pfam families, our method CoinDCA significantly exceeds the others when the top L/10, L/5 and L/2 predicted contacts are evaluated, no matter whether the contacts are short-, medium- and long-range. When neither auxiliary families nor supervised learning is used, CoinDCA is exactly the same as PSICOV. Therefore, the results in Table 1 indicate that combining joint EC analysis and supervised learning indeed can improve contact prediction accuracy over single-family EC analysis. We also observe the following performance trends.
1) In terms of contact prediction, the MSAs generated by PSIBLAST or HHblits are not better than the MSAs taken directly from the Pfam database.
2) A simple majority voting even underperforms the single-family EC method. This may be due to a couple of reasons. When a single family is considered, PSICOV may make wrong prediction for each family in very different ways due to the huge search space, so consensus of single-family results can only identify those highly-conserved contacts, but not those specific to one or few families. In addition, majority voting may suffer from alignment errors.
3) It does not work well by merging the target and auxiliary families into a single MSA and then applying single-family EC analysis. There are two possible reasons. One is that the resultant MSA may contain alignment errors, especially when the auxiliary families are not very close to the target family. The other is that we cannot use a single multivariate Gaussian distribution to model the related but different families due to sequence divergence (at some positions). In particular, Merge_p (i.e., merging by a probabilistic method) performs better than Merge_m (i.e., merging by MCoffee), so we will use Merge_p as the representative family merging method.

PSICOV models the MSA of a protein family using a multivariate Gaussian distribution. This Gaussian assumption holds only when the family contains a large number of sequence homologs. PlmDCA and GREMLIN do not use the Gaussian assumption and are reported to outperform PSICOV on some Pfam families (Kamisetty, et al., 2013). Our method CoinDCA still uses the Gaussian assumption. This test result indicates that when EC information in multiple related families is used, even with Gaussian assumption, we can still outperform those single-family EC methods without Gaussian assumption.

**Table 1**. Contact prediction accuracy on all the 98 test Pfam families. plmDCA and GREMLIN use the MSAs in the Pfam database while plmDCA_h and GREMLIN_h use the MSAs generated by HHblits.

|  | Short-range | | | Medium-range | | | Long-range | | |
| --- | --- | --- | --- | --- | --- | --- | --- | --- | --- |
|  | L/10 | L/5 | L/2 | L/10 | L/5 | L/2 | L/10 | L/5 | L/2 |
| **CoinDCA** | **0.528** | **0.446** | **0.316** | **0.496** | **0.435** | **0.312** | **0.561** | **0.502** | **0.391** |
| PSICOV | 0.369 | 0.299 | 0.205 | 0.375 | 0.312 | 0.213 | 0.446 | 0.400 | 0.311 |
| PSICOV_b | 0.356 | 0.286 | 0.199 | 0.388 | 0.306 | 0.199 | 0.462 | 0.400 | 0.294 |
| Merge_p | 0.316 | 0.265 | 0.183 | 0.303 | 0.246 | 0.178 | 0.370 | 0.328 | 0.253 |
| Merge_m | 0.298 | 0.237 | 0.172 | 0.276 | 0.223 | 0.169 | 0.355 | 0.309 | 0.232 |
| Voting | 0.343 | 0.232 | 0.184 | 0.405 | 0.280 | 0.168 | 0.337 | 0.353 | 0.275 |
| plmDCA | 0.422 | 0.327 | 0.203 | 0.433 | 0.354 | 0.233 | 0.484 | 0.443 | 0.343 |
| plmDCA_h | 0.387 | 0.300 | 0.186 | 0.433 | 0.339 | 0.211 | 0.480 | 0.413 | 0.292 |
| GREMLIN | 0.410 | 0.312 | 0.220 | 0.401 | 0.332 | 0.225 | 0.447 | 0.423 | 0.329 |
| GREMLIN_h | 0.387 | 0.291 | 0.188 | 0.391 | 0.316 | 0.204 | 0.428 | 0.400 | 0.301 |

## Performance with respect to the number of sequence homologs

Our method outperforms the others regardless of the size of a protein family. Similar to (Marks, et al., 2011; Wang and Xu, 2013), we calculate the number of non-redundant sequence homologs in a family (or multiple sequence alignment) by $M_{eff} = \sum_i \frac{1}{\sum_j s_{i,j}}$ where $i$ and $j$ are sequence indexes and $s_{i,j}$ is a binary variable indicating if two sequences are similar or not. It is equal to 1 if the normalized hamming distance between two sequences is less than 0.3; otherwise, 0. We divide the 98 test families into 5 groups by $\ln M_{eff}$:[4,5), [5,6), [6,7), [7,8), [8,10) and calculate the average L/10 prediction accuracy in each group. Figure 1 shows that our method performs significantly better than the others regardless of $\ln M_{eff}$. In particular, the advantage of our method over the others is even larger when $\ln M_{eff}$ is small.

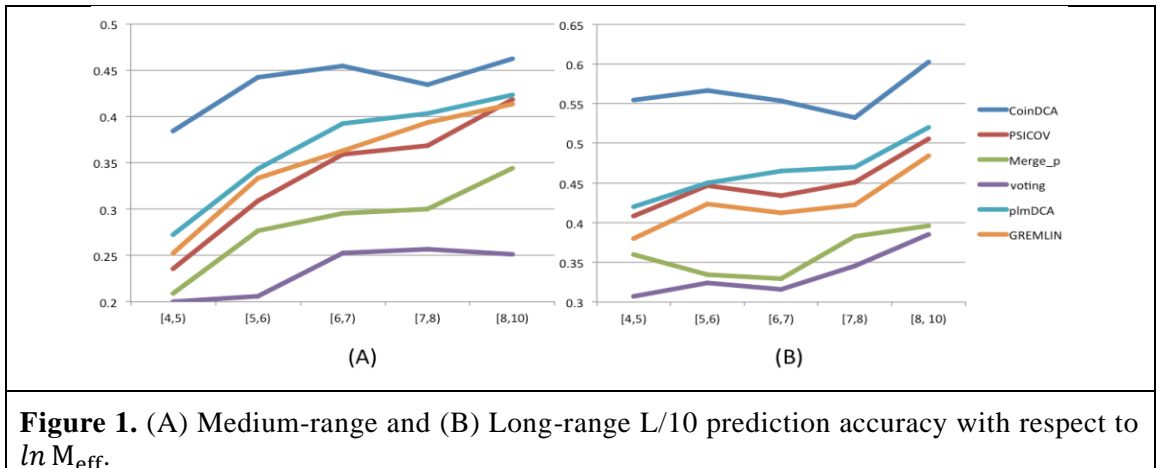

**Figure 1.** (A) Medium-range and (B) Long-range L/10 prediction accuracy with respect to $\ln M_{eff}$.

## Performance with respect to contact conservation level

For a native contact in the target family, we measure its conservation level by the number of its aligned native contacts in the auxiliary families. The 98 test families have conservation levels ranging from 0 to 8, corresponding to non-conserved and highly-conserved, respectively. In particular, a native contact with a conservation level of 0 is target family-specific since it has no support from any auxiliary families. Correct prediction of family-specific contacts is important since they may be very useful to the refinement of a template-based protein model.

Figure S1 in Appendix shows the distribution of the contact conservation level in our test set. As shown in this figure, a large number of native contacts are not conserved. Figure 2 (A) and (B) shows the ratio of medium- and long-range native contacts ranked among top $L/10$ predictions with respect to contact conservation level. Our method CoinDCA ranks more native long-range contacts among top $L/10$ than the single-family EC methods PSICOV, plmDCA and GREMLIN regardless of conservation level. CoinDCA has similar performance as the family merging method Merge_p for long-range contacts with conservation level ≥5, but significantly outperforms Merge_p for family-specific contacts. This may be because when the target and auxiliary families are merged together, the signal for highly-conserved contacts is reinforced but that for family-specific contacts is diluted. By contrast, our joint EC analysis method can reinforce the signal for highly-conserved contacts without losing family-specific information.

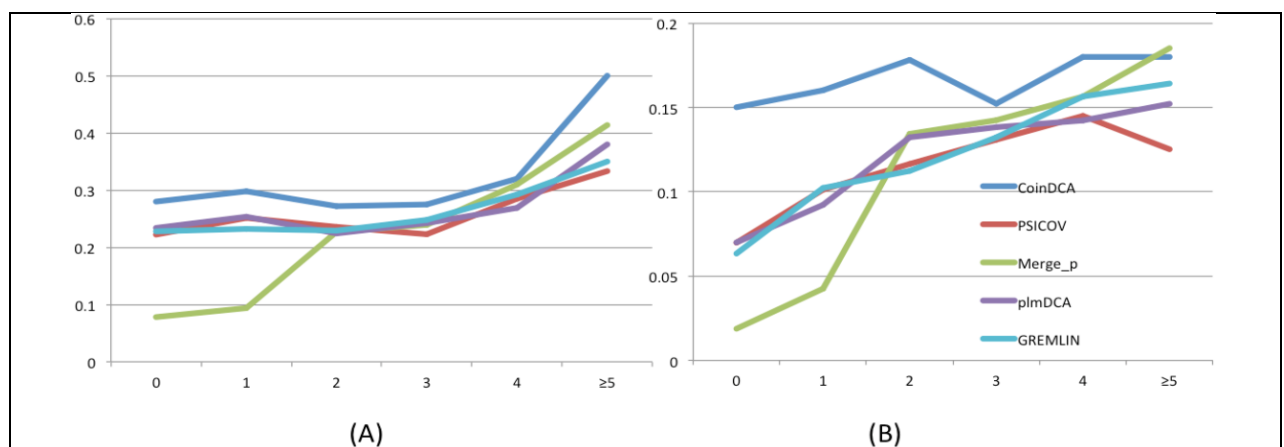

**Figure 2.** Prediction accuracy with respect to contact conservation level for (A) medium-range and (B) long-range. The Y-axis is the ratio of native contacts ranked among top $L/10$ predictions by a prediction method.

## Performance on the CASP10 test set

We further evaluate our method on the CASP10 targets with two supervised learning methods NNcon and CMAPpro and four EC analysis methods PSICOV, plmDCA, GREMLIN and Evfold. We run NNcon, PSICOV, plmDCA, GREMLIN and Evfold locally with default parameters, and CMAPpro through its web server. Again, we run HHpred to search the Pfam database for auxiliary families for the CASP10 targets. Meanwhile, 75 of 123 targets have at least one auxiliary family. For those targets without any auxiliary families, our method actually becomes the combination of single-family EC analysis and supervised learning. As shown in Table 2, on the whole CASP10 set, our method CoinDCA significantly exceeds the others in terms of the accuracy of the top $L/10$, $L/5$ and $L/2$ predicted contacts.

**Table 2**. Contact prediction accuracy on all the 123 CASP10 targets.

|  | Short-range | | | Medium-range | | | Long-range | | |
|---|---|---|---|---|---|---|---|---|---|
|  | L/10 | L/5 | L/2 | L/10 | L/5 | L/2 | L/10 | L/5 | L/2 |
| **CoinDCA** | **0.517** | **0.435** | **0.311** | **0.500** | **0.440** | **0.340** | **0.412** | **0.351** | **0.279** |
| PSICOV | 0.234 | 0.191 | 0.140 | 0.310 | 0.259 | 0.192 | 0.276 | 0.225 | 0.168 |
| plmDCA | 0.264 | 0.218 | 0.152 | 0.344 | 0.289 | 0.214 | 0.326 | 0.280 | 0.213 |
| NNcon | 0.499 | 0.399 | 0.275 | 0.393 | 0.334 | 0.226 | 0.239 | 0.188 | 0.001 |
| GREMLIN | 0.256 | 0.212 | 0.161 | 0.343 | 0.280 | 0.229 | 0.320 | 0.278 | 0.159 |
| CMAPpro | 0.437 | 0.368 | 0.253 | 0.414 | 0.363 | 0.276 | 0.336 | 0.297 | 0.227 |
| Evfold | 0.193 | 0.165 | 0.130 | 0.294 | 0.249 | 0.188 | 0.257 | 0.225 | 0.171 |

We also divide the 123 CASP10 targets into five groups according to $\ln M_{eff}$: (0,2), (2,4), (4,6), (6,8), (8,10), which contain 19, 17, 25, 36 and 26 targets, respectively. Meanwhile, $M_{eff}$ measures the number of non-redundant sequence homologs available for a target protein under prediction. We then calculate the average medium- and long-range contact prediction accuracy in each group. Figure 3 clearly shows that the prediction accuracy increases with respect to $\ln M_{eff}$ and that our method outperforms the others regardless of $\ln M_{eff}$. In particular, our method works much better than the single-family EC analysis methods when $\ln M_{eff} < 8$.

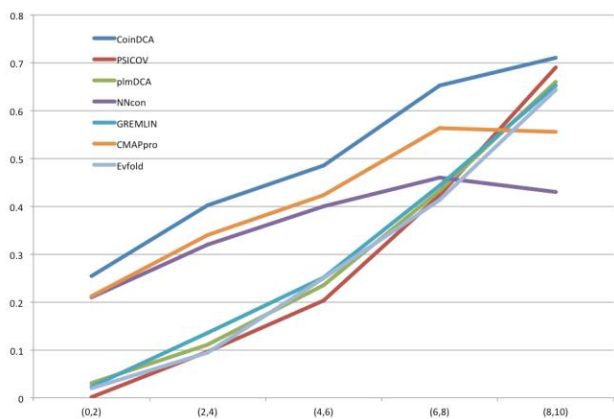

**Figure 3.** The relationship between prediction accuracy and $\ln M_{eff}$. X-axis is the $\ln M_{eff}$ value and Y-axis is the mean accuracy of top L/10 predicted contacts in the corresponding CASP10 target group. Only medium- and long-range contacts are considered.

## Discussion

This paper has presented a GGL method to predict contacts by integrating joint EC analysis and supervised machine learning. Evolutionary coupling (EC) analysis and supervised learning are currently two major methods for contact prediction, but they use different types of information. Our joint EC analysis predicts contacts in a target family by analyzing residue co-evolution information in a set of related protein families which may share similar contact maps. In order to effectively integrate information across multiple families, we use GGL to estimate the joint probability distribution of multiple related families by a set of correlated multivariate Gaussian distributions. Experiments show that the combination of joint EC analysis with supervised machine learning can significantly improve contact prediction, and that our method greatly improves contact prediction over single-family EC analysis even for protein families with a large number of sequence homologs. We have also shown that contact prediction cannot be improved by consensus of single-family EC analysis results, and that a family merging methods can improve prediction for highly-conserved contacts at the cost of family-specific contacts. This paper shows that.

Our method can be further improved. For example, similar to GREMLIN (Kamisetty, et al., 2013) and plmDCA (Ekeberg, et al., 2013), we can relax the Gaussian assumption to improve prediction accuracy. This paper uses an entry-wise $L_2$ norm to penalize contact map inconsistency among related protein families. There may be other penalty functions that can more accurately quantify contact map similarity between two families as a function of sequence similarity and thus, further improve contact prediction. There may be other better methods to integrate the predicted contact probability (by supervised learning) into our GGL framework. It may further improve contact prediction by integrating other supervised learning methods such as CMAPpro, NNcon and DNcon into our GGL framework.

In this paper we use Pfam to define a protein family because it is manually-curated and very accurate. There are also other criteria to define a protein family. For example, SCOP defines a protein family based upon structure information and thus, classifies protein domains into much fewer families than Pfam. In our experiment, the average structure similarity, measured by TMscore (Zhang and Skolnick, 2005), between a target (Pfam) family and its auxiliary (Pfam) families is only around 0.7. That is, many auxiliary families are not highly similar to its target families even by the SCOP definition. Indeed, some auxiliary families are only similar to the target family at the SCOP fold level. That is, even a remotely-related protein family may provide information useful for contact prediction.

We can further extend our method to predict contacts of all the protein families simultaneously, instead of one-by-one, by joint EC analysis across the whole protein family universe. First we can use a graph to model the whole Pfam database, each vertex representing one Pfam family and an edge indicating that two families may be related. Then we can use a graph of correlated multivariate Gaussian distributions to model the whole Pfam graph, each distribution for one vertex. The distributions of two vertices in an edge are correlated together through the alignment of their respective protein families. By this way, the residue co-evolution information in one family can be passed onto any family that is connected through a path. As such, we may predict the contacts of one family by making use of information in all the path-connected families. By enforcing this global consistency, we shall be able to further improve EC analysis for contact prediction. However, to simultaneously estimate the parameters of all the Gaussian distributions, we will need a large amount of computational power.

## Methods

### Probabilistic model of a single protein family

Modeling a single protein family using a probabilistic graphical model has been described in a few papers (Balakrishnan, et al., 2011; Jones, et al., 2012; Ma, et al., 2014; Marks, et al., 2011). Here we briefly introduce it since it is needed to understand our joint graphical model. Given a protein family $k$ and the MSA (multiple sequence alignment) of its sequences, let $X$ denote this MSA where $X_{ir}^k$ is a 21-dimension binary vector indicating the amino acid type (or gap) at row r (of this MSA) and column $i$ and $X_{ir}^k(a)$ is equal to 1 if the amino acid at row r (of this MSA) and column $i$ is $a$. Let $\bar{X}_i^k$ denote the mean vector of $X_{ir}^k$ across all the rows (i.e., proteins). Let $L$ denote the sequence length of this family and $N_k$ the number of sequences. Assuming this MSA has a Gaussian distribution $N(\mu^k, \Sigma^k)$ where $\mu^k$ is the mean vector with $21L$ elements and $\Sigma^k$ the covariance matrix of size $21L \times 21L$. The covariance matrix consists of $L^2$ submatrices, each having size $21 \times 21$ and corresponding to two columns in the MSA. Let $\Sigma_{ij}^k$ denote the submatrix for columns i and j. For any two amino acids (or gap) $a$ and $b$, their corresponding entry $\Sigma_{ij}^k(a,b)$ can be calculated as follows.

$$\Sigma_{ij}^k(a,b) = \frac{1}{N_k}\sum_{r=1}^{N_k}(X_{ir}^k(a) - \bar{X}_i^k(a))(X_{jr}^k(b) - \bar{X}_j^k(b)) \qquad (1)$$

The $\Sigma^k$ calculated by Eq. (1) actually is an empirical covariance matrix, which can be treated as an estimation of the true covariance matrix. Let $\Omega^k = (\Sigma^k)^{-1}$ denote the inverse covariance matrix (also called precision matrix), which indicates the residue or column interaction (or co-evolution) pattern in this protein family. In particular, the zero pattern in $\Omega^k$ represents the conditional independence of the MSA columns. Similar to $\Sigma_{ij}^k$, the precision submatrix $\Omega_{ij}^k$ indicates the interaction strength (or inter-dependency) between two columns $i$ and $j$, which are totally independent (given all the other columns) if only if $\Omega_{ij}^k$ is zero.

Due to matrix singularity, we cannot directly calculate $\Omega^k$ as the inverse of the empirical covariance matrix. Instead, we may estimate $\Omega^k$ by maximum-likelihood with a regularization factor $\lambda_1$ as follows.

$$\max_{\Omega^k} \log P(X^k|\Omega^k) - \lambda_1 \|\Omega^k\|_1$$

Where $\|\Omega^k\|_1$ is the L$_1$ norm of $\Omega^k$, which is used to make $\Omega^k$ sparse. Since $P$ is Gaussian, the above optimization problem is equivalent to the following.

$$\max_{\Omega^k} \left(\log|\Omega^k| - \operatorname{tr}(\Omega^k \hat{\Sigma}^k)\right) - \lambda_1 \|\Omega^k\|_1$$

Where $\hat{\Sigma}^k$ is the empirical covariance matrix calculated from the MSA. The PSICOV method for contact prediction is based upon the above formulation.

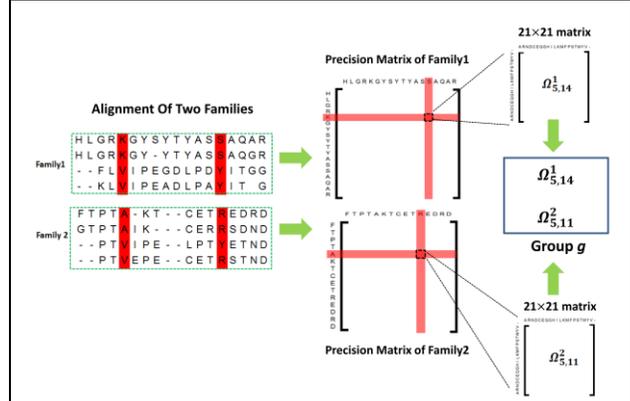

**Figure 4.** Illustration of column pair and precision submatrix grouping. Columns 5 and 14 in the 1$^{st}$ family are aligned to columns 5 and 11 in the 2$^{nd}$ family, respectively, so column pair (5,14) in the 1$^{st}$ family and the pair (5,11) in the 2$^{nd}$ family are assigned to the same group. Accordingly, the two precision submatrices $\Omega^1_{5,14}$ and $\Omega^2_{5,11}$ belong to the same group.

## Probabilistic model of multiple related protein families by Group Graphical Lasso (GGL)

The previous section introduces how to model a single protein family using a Gaussian graphical model (GGM). In this section we present our probabilistic model for a set of $K$ related protein families using a set of correlated GGMs. Here we still assume that each protein family has a Gaussian distribution with a precision matrix $\Omega^k$ ($k = 1, 2, \ldots, K$). Let $\Omega$ denote the set $\{\Omega^1, \Omega^2, \ldots, \Omega^K\}$ and $X = \{X^1, X^2, \ldots, X^K\}$ denote the set of MSAs. If we assume that the $K$ families are independent of each other, we can estimate their precision matrices by maximizing their joint log-likelihood as follows.

$$\max_{\Omega} \log P(X|\Omega) = \log \prod_{k=1}^{K} P(X^k|\Omega^k) - \lambda_1 \sum_{k=1}^{K} \|\Omega^k\|_1$$

$$= \sum_{k=1}^{K} \left(\log|\Omega^k| - \operatorname{tr}(\Omega^k \hat{\Sigma}^k)\right) - \lambda_1 \sum_{k=1}^{K} \|\Omega^k\|_1 \quad (2)$$

To model the correlation of these families, we assume that the precision matrices are correlated. Now we will show how to model the correlation of the precision matrices through the alignment of these protein families.

We build a multiple sequence alignment (MSA) of these $K$ protein families using a sequence alignment method. Each column in this MSA may consist of columns from several families. If column pair $(j_1, j_3)$ in family $k_1$ is aligned to column pair $(j_2, j_4)$, the interaction strength between two columns $j_1$ and $j_3$ in family $k_1$ shall be similar to that between columns $j_2$ and $j_4$ in family $k_2$. That is, if there is one contact between two columns $j_1$ and $j_3$, then it is very likely there is also a contact between two columns $j_2$ and $j_4$. Accordingly, the precision submatrix $\Omega^{k_1}_{j_1,j_3}$ for the two columns $j_1$ and $j_3$ in the family $k_1$ shall be related to the submatrix for the two columns $j_2$ and $j_4$ in the family $k_2$ (i.e., $\Omega^{k_2}_{j_2,j_4}$). The correlation strength between $\Omega^{k_1}_{j_1,j_3}$ and $\Omega^{k_2}_{j_2,j_4}$ depends on the conservation level of these two column pairs. That is, if these two column pairs are highly conserved, $\Omega^{k_1}_{j_1,j_3}$ and $\Omega^{k_2}_{j_2,j_4}$ shall also be highly correlated. Otherwise, they may be only weakly related. Based upon this observation, we divide all the column pairs into groups so that any two aligned column pairs belong to the same group, as shown in Figure 4. Therefore, if a target family has $L$ columns aligned to at least one auxiliary family, then there are in total $L(L-1)/2$ groups.

Let $G$ denote the number of groups and $K$ the number of involved families. We estimate the $K$ precision matrices by taking into account their correlation using group graphical lasso (GGL) as follows.

$$\max \sum_{k=1}^{K}\left(\log|\Omega^k| - \text{tr}(\Omega^k \hat{\Sigma}^k)\right) - \lambda_1 \sum_{k=1}^{K}\|\Omega^k\|_1 - \sum_{g=1}^{G} \lambda_g \|\Omega_g\|_2 \quad (3)$$

Where $g$ represents one group and $\|\Omega_g\|_2 = \sqrt{\sum_{(i,j,k)\in g}\|\Omega_{i,j}^k\|_F^2}$ where $\|\Omega_{i,j}^k\|_F^2$ is the square of the entry-wise $L_2$ norm of the precision submatrix $\Omega_{i,j}^k$. By using this penalty item, we ensure that the column pairs in the same group have similar interaction strength. That is, if one column pair in a particular group has a relatively strong interaction (i.e., $\|\Omega_{i,j}^k\|_F^2$ is large), the other column pairs in this group shall also have a larger interaction strength. In the opposite, if one column pair in a particular group has a relatively weak interaction (i.e., $\|\Omega_{i,j}^k\|_F^2$ is small), the other column pairs in this group shall also have a smaller interaction strength.

The parameter $\lambda_g$ is used to enforce residue co-evolution consistency in the same group. It is proportional to the conservation level in group $g$. We measure the conservation level using both the square root of the number of aligned families in a group and also the alignment probabilities. In particular, $\lambda_g$ is defined as follows.

$$\lambda_g = \alpha \sqrt{N-1} \sqrt[N-1]{\prod_{n=1}^{N-1} P_n}$$

Where $\alpha$ is a constant (=0.001), $N$ is the number of column pairs in group $g$ and $P_n$ can be interpreted as the average alignment probability between the target family and the auxiliary family $n$ at the two aligned columns belonging to group $g$. Meanwhile, $P_n$ is calculated as $P_n = P_i P_j$ where $P_i$ and $P_j$ are the marginal alignment probabilities at the two aligned columns. That is, when the two aligned column pairs are conserved, both $P_i$ and $P_j$ are large, so is $P_n$. Consequently, $\lambda_g$ is large and thus the interaction strength consistency among the column pairs in group $g$ is strongly enforced. In the opposite, if the marginal alignment probability is relatively small, $\lambda_g$ is small. In this case, we shall not strongly enforce the interaction strength consistency among column pairs in this group. By using the conservation level (or alignment quality) to control the correlation of interaction strength, our method is robust to bad alignments and thus, can also deal with protein families similar at different levels.

Note that our formulation (3) differs from the PSICOV formulation only in the last term, which is used to enforce co-evolution pattern consistency among multiple families. Without this term, our formulation is exactly the same as PSICOV when the same $\lambda_1$ is used. We use an ADMM (Hestenes, 1969) algorithm to solve formulation (3), which is described in Appendix.

### Including predicted probability by supervised learning as prior information

Compared to single-family EC analysis, our joint EC analysis uses residue co-evolution information from auxiliary families to improve contact prediction. In addition to co-evolution information, sequence profile and some non-evolutionary information are also useful for contact prediction. To make use of them, we first use a supervised machine learning method Random Forests to predict the probability of two residues forming a contact and then integrate this predicted probability as prior into our GGL framework. In particular, our Random Forests model predicts the probability of two residues forming a contact using the following information.
1) PSI-BLAST sequence profile. To predict the contact probability of two residues, we use their position-specific mutation scores and those of the sequentially adjacent residues.
2) Mutual information (MI) and its power series. When residue A has strong interaction with B and B has strong interaction with residue C, it is likely that residue A also has interaction with C. We use the MI power series to account for this kind of chaining effect. In particular, we use $MI^k$ where $k$ ranges from 2 to 11 where MI is the mutual information matrix. When there are many sequence homologs, the MI power series are very helpful to medium- and long-range contact prediction.
3) Non-evolutionary information such as residue contact potential described in (Tan, et al., 2006).
4) **EPAD:** a context-specific distance-dependent statistical potential (Zhao and Xu, 2012), derived from protein evolutionary information. The $C_\alpha$ and $C_\beta$ atomic interaction potential at all the distance bins is used. The atomic distance is discretized into bins by 1Å and all the distance>15Å is grouped into a single bin.
5) Homologous pairwise contact score. This score quantify the probability of a residue pair forming a contact between two secondary structure types. See paper (Wang and Xu, 2013) for more details.
6) Amino acid physic-chemical properties.

Some features are calculated on the residues in a local window of size 5 centered at the residues under consideration. In total there are ~300 features for each residue pair.

We trained and selected the model parameters of our Random Forests model by 5-fold cross validation. In total we used about 850 training proteins, all of which have <25% sequence identity with our test proteins. See paper (Wang and Xu, 2013) for the description of the training proteins.

Finally, our GGL formulation with predicted contact probability as prior is as follows.

$$\max \sum_{k=1}^{K} \left( \log|\Omega^k| - \text{tr}(\Omega^k \hat{\Sigma}^k) \right) - \lambda_1 \sum_{k=1}^{K} \|\Omega^k\|_1 - \sum_{g=1}^{G} \lambda_g \|\Omega_g\|_2 - \lambda_2 \sum_{k=1}^{K} \sum_{ij} \frac{\|\Omega_{ij}^k\|_1}{max(P_{ij}^k, 0.3)} \quad (4)$$

Where $P_{ij}^k$ is the predicted contact probability by Random Forests and $max(P_{ij}^k, 0.3)$ is used to reduce the impact of very small predicted probability. Meanwhile, $\exp(-\lambda_2 \sum_{k=1}^{K} \sum_{ij} \frac{\|\Omega_{ij}^k\|_1}{max(P_{ij}^k, 0.3)})$ can be interpreted as the prior probability of $\Omega$, which is used to promote the similarity between the precision matrix and the predicted contact probability. Formulation (4) differs from formulation (3) only in the last term. From computational perspective, $\lambda_2 \sum_{k=1}^{K} \sum_{ij} \frac{\|\Omega_{ij}^k\|_1}{max(P_{ij}^k, 0.3)}$ is similar to $\lambda_1 \sum_{k=1}^{K} \|\Omega^k\|_1$, so we can use almost the same computational method to optimize both formulations.

## Alignment of multiple protein families
To build the alignment of multiple protein families, we employ a probabilistic consistency method in (Do, et al., 2006; Peng and Xu, 2011). To employ this consistency method, we need to calculate the probabilistic alignment matrix between any two protein families. Each matrix entry is the marginal alignment probability (MAP) of two columns, each in one family. In addition to this probability method, we also employed MCoffee (Wallace, et al., 2006) to generate alignment of multiple families.

## Majority voting method for contact prediction
Majority voting is a simple way of utilizing auxiliary protein families for contact prediction. We first build an alignment of multiple protein families using the methods mentioned above. Then we use PSICOV to predict contact map for each of the related protein families. To determine if there is a contact between any two columns $i$ and $j$ in the target protein family, we use a majority voting based upon the predicted contacts for all the column pairs aligned to the pair $(i, j)$. In addition, we also assign a weight to each family proportional to the number of non-redundant sequence homologs in it. The more NR sequence homologs, the more weight this family carries since usually such a family has higher contact prediction accuracy. In this experiment, each protein family is modeled using a different probability distribution since PSICOV is applied to each of the related families separately.

## Pre-processing and Post-processing

We employ the same pre- and post-processing procedures as PSICOV to ensure our comparison with PSICOV is fair. Briefly, to reduce the impact of redundant sequences, we apply the same sequence weighting method as PSICOV. In particular, duplicate sequences are removed and columns containing more than 90% of gaps are also deleted. The sequence is weighted using a threshold of 62% sequence identity. We add a small constant (=0.1) to the diagonal of the empirical covariance matrix to ensure it is not singular. Similar to PSICOV and plmDCA (Ekeberg, et al., 2013), average-product correction (APC) (Dunn, et al., 2008) is applied to post-process predicted contacts.

## Appendix

### Estimating precision matrices by Alternating Directions Method of Multipliers

Since the same algorithm can be used to optimize Eqs. (3) and (4), for simplicity we explain how to optimize Eq (3) using Alternating Directions Method of Multipliers (ADMM). See (Hestenes, 1969) for a tutorial of ADMM. To estimate the precision matrix for the target protein family, we shall solve the following optimization problem.

(P1) $\quad \max_\Omega f(\Omega) - \lambda_1 \sum_{k=1}^{K} ||\Omega^k||_1 - \lambda_g \sum_{g=1}^{G} ||\Omega_g||_2$ (5)

$$f(\Omega) = \sum_{k=1}^{K} \left( \log|\Omega^k| - \text{tr}(\Omega^k \hat{\Sigma}^k) \right)$$

To apply ADMM, we start from rewriting P1 as a constraint optimization problem by making a copy of $\Omega$ to $Z$, but without changing the solution space.

(P2) $\quad \max_{\Omega,Z} f(\Omega) - P(Z)$ (6)

$$\forall k, \ \Omega^k = Z^k, \quad P(Z) = \lambda_1 \sum_{k=1}^{K} ||Z^k||_1 + \lambda_g \sum_{g=1}^{G} ||Z_g||_2$$

Where $Z$ denote the set $\{Z^1, Z^2, \ldots, Z^K\}$. Eq. (6) can be augmented by adding one term to penalize the difference between $\Omega^k$ and $Z^k$ as follows.

(P3) $\quad \max_{\Omega,Z} f(\Omega) - P(Z) - \sum_{k=1}^{K} \frac{\rho}{2} \left|\left| \Omega^k - Z^k \right|\right|_F^2$ (7)

$$\forall k, \ \Omega^k = Z^k$$

P3 is equivalent to P2 and P1, but converges faster due to the penalty term. Here $\rho$ is a hyper-parameter controlling the convergence rate. Some heuristics methods were proposed for choosing $\rho$, such as (Boyd, et al., 2011; Wahlberg, et al., 2012). Empirically, setting $\rho$ to a constant (=0.1), our algorithm converges within 10 iterations for most cases. Using a Lagrange multiplier $U^k$ for each constraint $\Omega^k = Z^k$, we obtain the following Lagrangian dual problem.

(P4) $\quad \min_U \max_{\Omega,Z} f(\Omega) - P(Z) - \sum_{k=1}^{K} (\rho(U^k)^T (\Omega^k - Z^k) + \frac{\rho}{2} \left|\left| \Omega^k - Z^k \right|\right|_F^2)$ (8)

It is easy to prove that P3 is upper bounded by P4. We can solve P4 iteratively using a subgradient method. At each iteration, we fix the value of $U$ and solve the following subproblem.

(P5) $\quad \max_{\Omega,Z} f(\Omega) - P(Z) - \frac{\rho}{2} \sum_{k=1}^{K} ||\Omega^k - Z^k + U^k||_F^2$ (9)

The subgradient of $U$ is $-\rho(\Omega - Z)$, so we may update $U$ by $U + \rho(\Omega - Z)$ and repeat solving P5 until convergence, i.e., the difference between $\Omega$ and $Z$ is small.

To solve P5, we decompose it into the below two subproblems and then solve them alternatively.

(SP1) $\quad \forall k, \ (\Omega^k)^* = argmax\{f(\Omega) - \frac{\rho}{2} \sum_{k=1}^{K} ||\Omega^k - Z^k + U^k||_F^2\}$ (10)

(SP2) $\quad Z^* = argmin\{\frac{\rho}{2} \sum_{k=1}^{K} \left|\left| \Omega^k - Z^k + U^k \right|\right|_F^2 + P(Z)\}$ (11)

The sub-problem SP1 optimizes the objective function (9) with respect to $\Omega$ while fixing $Z$. Notice that in SP1 no two $\Omega^K$ are coupled together, so we can split it into $K$ independent optimization subproblems. The sub-problem SP2 optimizes the objective function with respect to $Z$ while fixing $\Omega$. Next we will show how to solve these two sub-problems efficiently.

**Solving SP1.** SP1 is a concave and smooth function so we can solve it by setting its derivate to zero as follows.

$$\left(\Omega^{k^{-1}} - \hat{\Sigma}^k\right) - \rho\left(\Omega^k - Z^k + U^k\right) = 0 \quad (12)$$

Let $M^k = \hat{\Sigma}^k - \rho Z^k + \rho U^k$, by Eq. (12) we have $M^k = \Omega^{k^{-1}} - \rho\Omega^k$. That is, $M^k$ has the same eigenvalues and eigenvectors as $\Omega^{k^{-1}} - \rho\Omega^k$. Consequently, $\Omega^k$ has the same eigenvectors as $M^k$. Let $\delta_i$ and $m_i$ denote the $i^{th}$ eigenvalues of matrix $\Omega^k$ and $M^k$, respectively. Then we have $m_i = \delta_i^{-1} - \rho\delta_i$, from which we can solve $\delta_i$ as follows.

$$\delta_i = \frac{-m_i + \sqrt{m_i^2 + 4\rho}}{2\rho} \quad (13)$$

Therefore we can reconstruct $\Omega^k$ from $\delta_i$ and the eigenvectors of $M^k$. The main computation time is consumed in calculating the eigenvalues and eigenvectors of $M^k$. Since $M^k$ is symmetric and sparse, we can permute its rows and columns to obtain a diagonal block matrix, which can be done within running time linear in the number of non-zero elements in $M^k$. Then we divide $M^k$ it into small submatrices and calculate their eigenvalues and eigenvectors separately.

**Solving SP2.** SP2 is a non-differentiable convex function and we can solve it by setting its sub-gradients to zero. That is, for each $k$, we have the following equation.

$$\rho(Z_{i,j}^k - A_{i,j}^k) + \lambda_1 \frac{Z_{i,j}^k}{\beta_g} + \lambda_g t_{i,j}^k = 0 \quad (14)$$

Where $A^k = \Omega^k + U^k$, $t_{i,j}^k$ is the derivative of $|Z_{i,j}^k|$, and $\beta_g = ||Z_g||_2$. Meanwhile, $t_{i,j}^k$ is equal to any value between -1 and 1 when $Z_{i,j}^k$ is 0 and otherwise, equal to $sign(Z_{i,j}^k)$. To solve a particular $Z_{i,j}^k$ based on Eq. (14), we need to know the value of $\beta_g$, which depends on all the $Z_{i,j}^k$ in group $g$. That is, we cannot solve these $K$ optimization problems independently.

Let $S(x, c) = max(x - c \times sign(x), 0)$. Eq. (14) can be written as follows.

$$\left(1 + \frac{\lambda_g}{\rho\beta_g}\right) Z_{i,j}^k = S(A_{i,j}^k, \frac{\lambda_1}{\rho}) \quad (15)$$

Squaring both sides of Eq. (15) and summing up over all $(i, j, k) \in g$, we have the following equation.

$$\sum_{(i,j,k) \in g} \left(1 + \frac{\lambda_g}{\rho\beta_g}\right)^2 (Z_{i,j}^k)^2 = \sum_{(i,j,k) \in g} S(A_{i,j}^k, \frac{\lambda_1}{\rho})^2 \quad (16)$$

By definition, $\sum_{(i,j,k) \in g} (Z_{i,j}^k)^2 = \beta_g^2$. Since $\left(1 + \frac{\lambda_g}{\rho\beta_g}\right)^2$ is independent of $(i, j, k)$, the left hand side of Eq. (16) is equal to $\left(\beta_g + \frac{\lambda_g}{\rho}\right)^2$. Therefore, we can represent $\beta_g$ as follows.

$$\beta_g = \sqrt{\sum_{(i,j,k) \in g} S(A_{i,j}^k, \frac{\lambda_1}{\rho})^2} - \frac{\lambda_g}{\rho} \quad (17)$$

Plugging Eq. (17) back into Eq. (15), we obtain the value of $Z_{i,j}^k$ as follows.

$$Z_{i,j}^k = S(A_{i,j}^k, \frac{\lambda_1}{\rho})(1 - \frac{\lambda_g}{\rho\sqrt{\sum_{(i,j,k) \in g} S(A_{i,j}^k, \frac{\lambda_1}{\rho})^2}}) \quad (18)$$

## Distribution of contact conservation level

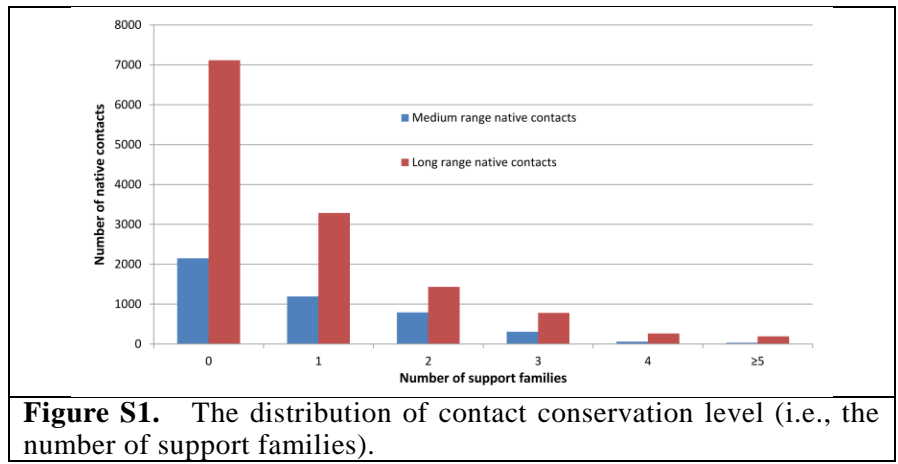

**Figure S1.** The distribution of contact conservation level (i.e., the number of support families).